\documentclass[10pt,twoside]{epnt1p}

\newcommand{\lsim}{\mathrel{\vcenter{\hbox{$<$}\nointerlineskip\hbox{$\sim$}}}} 

\newcommand{\egzk}{E_{\rm GZK}}
\newcommand{\rarr}{\rightarrow}
\newcommand{\sig}{\sigma^{\rm CC}_{\nu N}}

\newcommand{\Esh}{E^{\rm sh}_{\rm th}}
\newcommand{\Eth}{E^\tau_{\rm th}}

\newcommand{\lmin}{l_{\rm min}}

\newcommand{\dmin}{d_{\rm min}}
\newcommand{\dmax}{d_{\rm max}}

\newcommand{\zthin}{z_{\rm thin}}

\newcommand{\zc}{z_{\rm cloud}}

\newcommand{\numu}{\nu_\mu}
\newcommand{\nutau}{\nu_\tau}
\newcommand{\nue}{\nu_e}

\newcommand{\bwide}{\begin{widetext}}
\newcommand{\ewide}{\end{widetext}}
\newcommand{\beq}[1]{\begin{equation} \label{(#1)}}
\newcommand{\eeq}{\end{equation}}
\newcommand{\barr}{\begin{array}}
\newcommand{\earr}{\end{array}}
\newcommand{\ba}[1]{\begin{eqnarray} \label{(#1)}}
\newcommand{\ea}{\end{eqnarray}}

\usepackage{epsfig}

\usepackage{amssymb}
\usepackage{amsmath}

\usepackage{url}

\begin{document}
{\hfill\normalsize\tt IPPP/06/87}

{\hfill\normalsize\tt DCPT/06/174}

\begin{frontmatter}
\vspace{-1mm}
\title{Inferring Neutrino Cross Sections Above $10^{19}$~eV}

\author[address1]{Sergio Palomares-Ruiz}

\address[address1]{Institute for Particle Physics Phenomenology,
  University of Durham, Durham DH1 3LE, UK}

\vspace{-5mm}
\begin{abstract}
Extremely high energy neutrinos propagating in the atmosphere or in
the Earth can originate horizontal or upgoing air-showers,
respectively. We calculate the acceptances (event rate/flux) for
detecting both types of events by fluorescence detectors, both
space-based as with the EUSO and OWL proposals, and ground-based, as
with Auger, HiRes and Telescope Array.  We depict them as a function
of the neutrino-nucleon cross section, $\sig$, and show that from the
ratio of these two classes of events, the inference of $\sig$ above
$10^{19}$~eV appears feasible, assuming that a neutrino flux exists at
these energies. Our semi-analytic calculation includes realistic
energy-losses for tau leptons and Earth-curvature effects. We also
consider constraints on shower development and identification and the
effects of a cloud layer. 

\end{abstract}

\end{frontmatter}

\section{\label{sec:intro} Introduction}
Above the Greisen-Zatsepin-Kuzmin (GZK) energy of $\egzk\sim 5\times
10^{19}$~eV~\cite{GZK} ultra-high energy neutrinos  are probably
the only propagating primaries. Moreover, in contrast to cosmic-rays,
they point back to their astrophysical sources carrying
information not accesible with other primaries. The detection of
ultra-high energy neutrinos also allows studies of the fundamental
properties of neutrinos themselves, as for instance the
neutrino-nucleon cross section at energies beyond the reach of our
terrestrial accelerators.

In this talk, based on Ref.~\cite{PRIW}, we study the potential for
cosmic-ray experiments designed to track ultra-high energy air-showers
by monitoring their fluorescence yield, to detect horizontal
air-showers (HAS) and upgoing air-showers (UAS) induced by a cosmic 
neutrino flux and show the ability of these experiments to infer the
neutrino-nucleon cross section, $\sig$, at energies above $10^{19}$~eV,
from the ratio of their UAS and HAS events. Such energies are orders
of magnitude beyond the energies accessible to man-made terrestrial
accelerators. From the point of view of QCD, such a cross section
measurement would be an interesting microscope into the world of
small-x parton evolution. Deviations from QCD-motivated
extrapolations~\cite{GQRS} could reduce the cross section due to
saturation effects~\cite{saturation} or enhance it by the existance of 
new physics thresholds~\cite{newphysics}.

In Ref.~\cite{KW} it was shown that by comparing the HAS and UAS event
rates the neutrino-nucleon cross section may be inferred. The
calculation of Ref.~\cite{KW} gave an approximate result for the 
dependence of the UAS event rates on the neutrino-nucleon
cross section. In this talk, following the results of Ref.~\cite{PRIW},
we improve upon Ref.~\cite{KW} in several ways, as we show below. On
the other hand, the prospects of inferring the neutrino-nucleon
cross section at neutrino telescopes such as IceCube or at the Auger
observatory have been studied in Ref.~\cite{otherdetectors}.

\section{\label{sec:airshower} Air-shower rates and constraints on
  shower-development}

Ultra-high energy neutrinos are expected to arise from the decay of
pions and subsequently muons produced in astrophysical
sources~\cite{pimuchain} (for the case of production from neutron 
decays see, eg, Ref.~\cite{antinubeam}). After propagating for many
oscillation lengths and due to the maximal mixing between $\numu$ and
$\nutau$ inferred from terrestrial oscillation experiments, all
flavors are populated. Thus, a detector optimized for $\nue$ or
$\numu$ or $\nutau$ can expect a measurable flux from cosmic neutrinos.

The weak nature of the neutrino-nucleon cross section means that
HAS begin low in the atmosphere, where the target is most dense, and
thus that the event rate for neutrino-induced HAS is proportional to
the cross section. Following Ref~\cite{PRIW}, for the case of HAS
event rates we will only consider $\nu_e$ charged current interactions.

For a neutrino-induced UAS, the dependence on the neutrino
cross section is more complicated. The Earth itself is opaque for
neutrinos with energies exceeding about a PeV of energy. However,
``Earth-skimming'' neutrinos, those with a short enough chord length
through the Earth, will penetrate and exit, or penetrate and
interact. In particular, there is much interest in the Earth-skimming
process $\nutau\rarr\tau$ in the shallow Earth, followed by $\tau$
decay in the atmosphere to produce an observable shower. In
Ref.~\cite{KW} it was shown that the rate for the Earth-skimming
process $\nutau\rarr\tau$ is {\sl inversely} proportional to
$\sig$. The inverse dependence of UAS rate on $\sig$ is broken by
the $\tau\rarr$~{\sl shower} process in the atmosphere. As the
cross section decreases, the allowed chord length in the Earth
increases, and the tau emerges with a larger angle from the Earth's
tangent plane. This in turn provides a smaller path-length in air in
which the tau may decay and the resulting shower may evolve. This
effect somewhat mitigates the inverse dependence of the UAS on
$\sig$. 

The main aim of the study in Ref.~\cite{PRIW} was to provide a
detailed and improved extension of the idea introduced in
Ref.~\cite{KW}. Hence, here we include the energy dependences of the
tau energy-losses in the Earth, and of the tau lifetime in the
atmosphere. For the energy-losses, we distinguish between tau
propagation in rock and in water. In the case of the UAS, the
pathlength of the pre-decayed tau may be so long that the Earth's
curvature represents a non-negligible correction, that we include. We
also consider the partial loss of visibility due to cloud layers. On
the issue of shower development, we incorporate the dependence of
atmospheric density on altitude and add some conditions for the
showers to be observable. Shower detection will require
that within the field of view, the length of the shower track
projected on a plane tangent to the Earth's surface exceeds some
minimum length, $\lmin$. In addition, a minimum column density,
$\dmin$, beyond the point of shower initiation is required for the
shower to develop in brightness. On the other hand, after a maximum
column density, $\dmax$, the shower particles are below threshold for
further excitation of the $N_2$ molecules which provide the observable
fluorescence signal. Therefore, visible showers end at
$\dmax$. Finally, the fluorescent emission per unit length of the
shower will decline exponentially with the air density at 
altitude. We will take $\zthin= 24$~km as the altitude beyond which
the signal becomes imperceptible. Regarding the choice of $\dmin$ and
$\dmax$, they are inferred from the observed longitudinal development
profiles of ultrahigh-energy cosmic ray showers. On the other hand, we
assign a relatively small value to $\lmin$ to maximize the observable
event rate. For a summary of the different values adopted to obtain
the results, full details on the analytic description of the effects
of these parameters on the event rates and comparison with prior work,
we defer the reader to Ref.~\cite{PRIW}.

\section{\label{sec:results} Results}
In this section, we present the results of our semi-analytical
approach and take the product of area and solid angle $\sim 10^6\,{\rm
  km}^2\;{\rm sr}$, i.\ e.\ that of the EUSO design report~\cite{EUSO}.

\begin{figure*}[t]
\includegraphics[width=1.\textwidth]{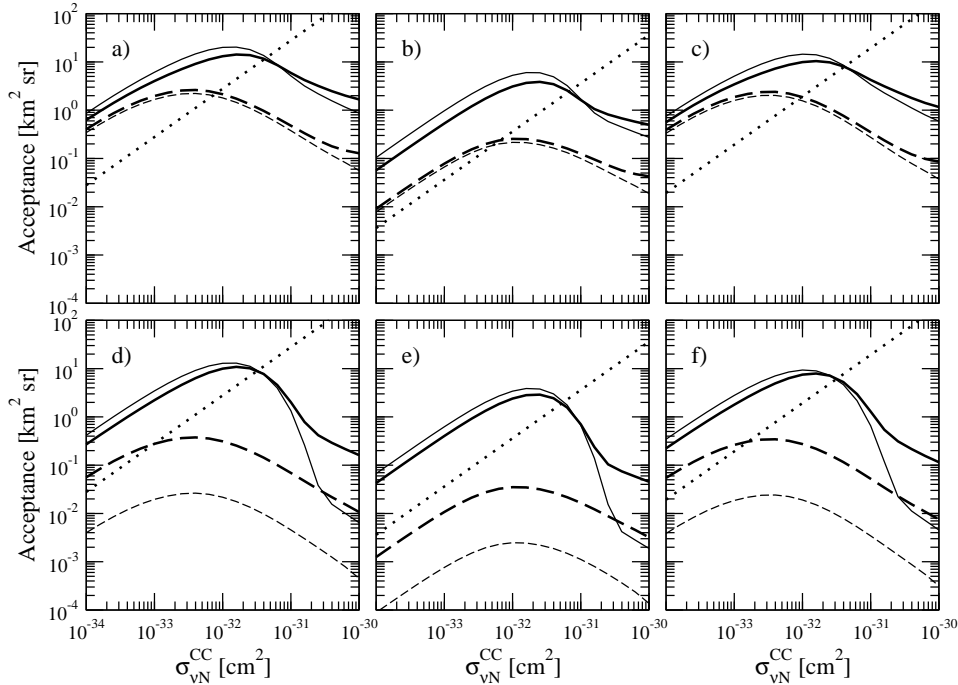}%
\caption{\label{combiEPNT} 
  Acceptances with $l_{\rm{min}}$ fixed at 5~km, $\dmin$ at $400 {\rm
  g}/{\rm cm}^2$, and $\dmax$ at $1200 {\rm g}/{\rm cm}^2$. The curves
  correspond to HAS (dotted line) and UAS over ocean with
  $E_\nu=10^{21}$~eV (thick solid line), ocean with $E_\nu=10^{20}$~eV
  (thin solid line), land with $E_\nu=10^{21}$~eV (thick dashed line),
  and land with $E_\nu=10^{20}$~eV (thin dashed line). Panels are for
(a) space-based (or ground-based) detectors in the absence of clouds
  with $\Esh=10^{19}$~eV;
(b) ground-based detectors in the presence of a cloud layer at
  $z_{\rm{cloud}} = 2$ km with $\Esh=10^{19}$~eV;
(c) spaced-based detectors in the presence of a cloud layer at
  $z_{\rm{cloud}} = 2$ km with $\Esh=10^{19}$~eV;
(d) space-based (or ground-based) detectors in the absence of clouds
  with $\Esh=5 \times 10^{19}$~eV;
(e) ground-based detectors in the presence of a cloud layer at
  $z_{\rm{cloud}} = 2$ km with $\Esh=5 \times 10^{19}$~eV;
(f) spaced-based detectors in the presence of a cloud layer at
  $z_{\rm{cloud}} = 2$ km with $\Esh=5 \times 10^{19}$~eV.
} 
\end{figure*}

In Fig.~\ref{combiEPNT} are plotted UAS (solid and dashed) and HAS
(dotted) acceptances in units of (${\rm km}^2$-sr), versus fixed
values of $\sig$. Within the approximations followed in
Ref.~\cite{PRIW}, for the ideal case of a cloudless sky (panels a and
d) there is no difference between the acceptances for ground-based and
space-based detectors. However, there are significant up-down
differences when the sky is covered by clouds (panels b, c, e and
f). In this latter case, we model the cloud layer as infinitely thin
with altitude $\zc$, but with an infinite optical depth so that
showers are completely hidden on the far side of the cloud layer. 

The HAS acceptances depend on neutrino energy only via $\sig (E_\nu)$,
and rise linearly with $\sig$. Plotted against fixed $\sig$, then, the
straight-line HAS curves (dotted) are universal curves valid for any
$E_\nu$ exceeding the trigger threshold $\Esh$. The UAS acceptances
have a complicated dependence on $E_\nu$; it arises from the energy
dependences of $\nu$ propagation in the Earth, tau propagation in the
Earth, and path-length of the tau in the atmosphere before it decays,
the latter also affecting the visible shower characteristics. We can
clearly see that the UAS acceptance (and so also the rate) is
typically an order of magnitude larger when neutrinos traverse a layer
of ocean water, compared to a trajectory where they only cross
rock. Thus, the UAS event rate is enhanced over the ocean relative to
over land. The value of this enhancement depends on the shower
threshold-energy $\Esh$ of the detector (upper versus lower panels)
and on the neutrino-nucleon cross section in a non-trivial
way. On the other hand, quantitatively, the ground-based acceptances
are quite reduced by the low-lying clouds, whereas the space-based
acceptances are not, as one would expect. The suppression of the
ground-based acceptance is most severe for small cross sections, for
which the tau leptons emerge more vertically and disappear into the
clouds before their eventual shower occurs and develops. Ground-based
UAS acceptances are reduced by up to an order of magnitude over water,
and even more over land. Ground-based HAS acceptances are reduced by
an order of magnitude. For space-based detectors, the UAS acceptance
is reduced little by clouds at 2~km. Larger neutrino cross sections
lead to more tangential tau-showers which may hide below a low-lying
cloud layer. We see that UAS reductions are a factor of 2 for the
larger cross sections shown, and less for the smaller values of
cross section.

We obtain benchmark event rates by multiplying our calculated
acceptances with a benchmark integrated flux of one neutrino per
$({\rm km}^2\,{\rm sr}\,{\rm yr})$. The result is a signal exceeding
an event per year for an acceptance exceeding a (${\rm km}^2$-sr).
Thus we see that this benchmark flux gives a HAS rate exceeding 1/yr
if $\sig$ exceeds $10^{-32}\,{\rm cm}^2$; and an UAS rate exceeding
1/yr over water for the whole cross section range with
$\Esh=10^{19}$~eV, and over land if $\sig \lsim 10^{-31}\,{\rm
  cm}^2$. When $\Esh$ is raised, however, the UAS signal over land is 
seriously compromised, while  UAS rates over the ocean are little
changed, HAS rates are unchanged, as long as $\Eth$ exceeds $E_\nu$.

We call attention to the fact that for UAS over both ocean and land,
there is a maximum in the UAS acceptance at cross section values $\sig
\sim (1-2) \times 10^{-32}~\rm{cm}^2$ and $\sig \sim (0.3-0.5) \times
10^{-32}~\rm{cm}^2$, respectively. For cross sections similar or
smaller than those at the maximum, the acceptance for UAS is larger 
than that for HAS; conversely, for cross sections above those at the
maximum, HAS events will dominate UAS events. The cross section value
at the maximum lies just below the extrapolation of the Standard Model
cross section, which for the two initial neutrino energies considered,
$10^{20}$~eV and $10^{21}$~eV, is $0.54 \times 10^{-31} \rm{cm}^2$ and
$1.2 \times 10^{-31} \rm{cm}^2$, respectively. If this extrapolation
is valid, then one would expect comparable acceptances (and event
rates) for UAS over water and for HAS. If the true cross section
exceeds the extrapolation, then HAS events will dominate UAS events;
if the true cross section is suppressed compared to the extrapolation,
then UAS events will dominate HAS events. Importantly, the very
different dependences on the cross section of the HAS and UAS
acceptances offers a practical method to measure $\sig$. One has
simply to exploit the ratio of UAS-to-HAS event rates. Furthermore,
the shape of the UAS acceptance with respect to $\sig$ establishes the
``no-lose theorem''~\cite{PRIW,KW}, which states that although a large
cross section is desirable to enhance the HAS rate, a smaller
cross section still provides a robust event sample due to the
contribution of UAS.  The latter is especially true over ocean.

\section{Conclusions}
In this talk we have presented a mostly analytic calculation of the
acceptances of space-based and ground-based fluorescence detectors of
air-showers at ultra-high energies. Included in the calculation are the
dependences of the acceptances on initial neutrino energy,
trigger-threshold for the shower energy, composition of Earth (surface 
rock or ocean water), and several shower parameters (the minimum and
maximum column densities for shower visibility, and the tangent length
of the shower). Also included in the calculation are suppression of the
acceptances by cloud layers and by the Earth's curvature. And most
importantly, also included are the dependences on the unknown
neutrino-nucleon cross section. The dependence is trivial and linear
for HAS, but nontrivial and nonlinear for UAS.

The merits of the analytic construction are two-fold: it offers an
intuitive understanding of each ingredient entering the calculation;
and it allows one to easily re-compute when different parameters are
varied. While a Monte Carlo approach may be simpler to implement, it
sacrifices some insight and efficiency.

The differing dependences of HAS and UAS on $\sig$ enable two very
positive conclusions: (1) the ``no-lose theorem'' is valid, i.\ e.\ 
that acceptances are robust for the combined HAS plus UAS signal
regardless of the cross section value; (2) and an inference of $\sig$
above $10^{19}$~eV is possible if HAS and UAS are both measured.

\vspace{-5mm}
\section*{\bf Acknowledgments}

{\small 
I would like to thank T.~Weiler for enlightning discussions and for a
fruitful collaboration.
}

\setcounter{section}{0}
\setcounter{subsection}{0}
\setcounter{figure}{0}
\setcounter{table}{0}
\newpage

\vspace{-5mm}
\begin{thebibliography}{00}
{\small

\bibitem{GZK}
  K.~Greisen,
  Phys.\ Rev.\ Lett.\  {\bf 16}, 748 (1966);
%
  G.~T.~Zatsepin and V.~A.~Kuzmin,
  JETP Lett.\  {\bf 4}, 78 (1966)
  [Pisma Zh.\ Eksp.\ Teor.\ Fiz.\  {\bf 4}, 114 (1966)].


\bibitem{PRIW}
  S.~Palomares-Ruiz, A.~Irimia and T.~J.~Weiler,
  Phys.\ Rev.\ D {\bf 73}, 083003 (2006).


\bibitem{GQRS}
  R.~Gandhi, C.~Quigg, M.~H.~Reno and I.~Sarcevic,
  Phys.\ Rev.\ D {\bf 58}, 093009 (1998).


\bibitem{saturation}
  L.~V.~Gribov, E.~M.~Levin and M.~G.~Ryskin,
  Phys.\ Rept.\  {\bf 100}, 1 (1983).


\bibitem{newphysics}
  G.~Domokos and S.~Nussinov,
  Phys.\ Lett.\ B {\bf 187}, 372 (1987);
%
  H.~Aoyama and H.~Goldberg,
  Phys.\ Lett.\ B {\bf 188}, 506 (1987);
%
  A.~Ringwald,
  Nucl.\ Phys.\ B {\bf 330}, 1 (1990);
%
  S.~Nussinov and R.~Shrock,
  Phys.\ Rev.\ D {\bf 59}, 105002 (1999);
%
  G.~Domokos and S.~Kovesi-Domokos,
  Phys.\ Rev.\ Lett.\  {\bf 82}, 1366 (1999);
%
  L.~Anchordoqui and H.~Goldberg,
  Phys.\ Rev.\ D {\bf 65}, 047502 (2002);
%
  E.~J.~Ahn, M.~Cavaglia and A.~V.~Olinto,
  Phys.\ Lett.\ B {\bf 551}, 1 (2003).


\bibitem{KW}
  A.~Kusenko and T.~J.~Weiler,
  Phys.\ Rev.\ Lett.\  {\bf 88}, 161101 (2002).


\bibitem{otherdetectors}
  D.~Hooper,
  Phys.\ Rev.\ D {\bf 65}, 097303 (2002);
%
  L.~A.~Anchordoqui, J.~L.~Feng, H.~Goldberg and A.~D.~Shapere,
  Phys.\ Rev.\ D {\bf 66}, 103002 (2002);
%
  L.~A.~Anchordoqui {\it et al.},
  JCAP {\bf 0506}, 013 (2005);
%
  L.~Anchordoqui, T.~Han, D.~Hooper and S.~Sarkar,
  Astropart.\ Phys.\  {\bf 25}, 14 (2006);
%
  L.~Anchordoqui and F.~Halzen,
  Annals Phys.\  {\bf 321}, 2660 (2006);
%
  S.~Hussain, D.~Marfatia, D.~W.~McKay and D.~Seckel,
  Phys.\ Rev.\ Lett.\  {\bf 97}, 161101 (2006);
%
  V.~Barger, P.~Huber and D.~Marfatia,
  Phys.\ Lett.\ B {\bf 642}, 333 (2006).


\bibitem{pimuchain} 
See eg, 
%
  J.~P.~Rachen and P.~Meszaros,
  Phys.\ Rev.\ D {\bf 58}, 123005 (1998);
%
  L.~A.~Anchordoqui, H.~Goldberg, F.~Halzen and T.~J.~Weiler,
  Phys.\ Lett.\ B {\bf 621}, 18 (2005);
%
  T.~Kashti and E.~Waxman,
  Phys.\ Rev.\ Lett.\  {\bf 95}, 181101 (2005).


\bibitem{antinubeam}
  L.~A.~Anchordoqui, H.~Goldberg, F.~Halzen and T.~J.~Weiler,
  Phys.\ Lett.\ B {\bf 593}, 42 (2004);
%
  L.~A.~Anchordoqui {\it et al.},
  arXiv:astro-ph/0611581.


\bibitem{EUSO}
  ESA and EUSO Team, 2000, \url{http://www.euso-mission.org}

}
\end{thebibliography}
\end{document}